\documentclass{eas}
\usepackage{graphicx}
\def\CN2{\mbox{$C_N^2 \ $}}
\def\CT2{\mbox{$C_T^2 \ $}}

\def\tauO{\mbox{$\tau_0 \ $}}
\def\thetaO{\mbox{$\theta_{0} \ $}}

\def\sigmaI2{\mbox{$\sigma ^{2}_{I} \ $}}

\begin{document}

\title{Optical Turbulence above the Internal Antarctic Plateau} 
\author{E. Masciadri}\address{INAF, Osservatorio Astrofisico di Arcetri, L.go E. Fermi 5, Florence, Italy - \email{masciadri@arcetri.astro.it}}
\author{F. Lascaux}\sameaddress{1}
\author{S. Hagelin}\sameaddress{1}
\author{J. Stoesz}\sameaddress{1} 
\author{P. Le Moigne}\address{Centre National de Recherches Meteorologiques (CNRM), Meteo France, 42, Av. G. Coriolis, Toulouse, France}
\author{J. Noilhan}\sameaddress{2}
\begin{abstract}
The internal antarctic plateau revealed in the last years to be a site with interesting potentialities for the astronomical applications due to the extreme dryness and low temperatures, the typical high altitude of the plateau,  the weak level of turbulence in the free atmosphere down to a just few tens of meters from the ground and the thin optical turbulence layer developed at the ground. The main goal of a site testing assessment above the internal antarctic plateau is to characterize the site  (optical turbulence and classical meteorological parameters) and to quantify which is the gain we might obtain with respect to equivalent astronomical observations done above mid-latitude sites to support plans for future astronomical facilities. Our group is involved, since a few years, in studies related to the assessment of this site for astronomical applications that  include the characterization of the meteorological parameters and optical turbulence provided by general circulation models as well as mesoscale atmospherical models and the quantification of the performances of Adaptive Optics (AO) systems. In this talk I will draw the status of art of this site assessment putting our studies in the context of the wide international site testing activity that has been done in Antarctica. I will focus on the site assessment relevant for astronomical applications to be done in the visible up to the near infrared ranges, i.e. those ranges for which the optical turbulence represents a perturbing element for the quality of the images and the AO techniques an efficient tool to correct these wavefront perturbations.
\end{abstract}
\maketitle

\section{Dome C challenges for astronomical applications}
\label{intro}

\begin{figure}
\begin{center}
\includegraphics[width=5cm,angle=-90]{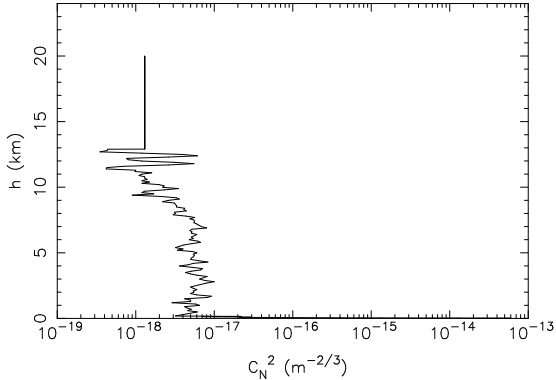}
\end{center}
\caption{Median value of the $\CN2$ measured in winter time above Dome C (from Trinquet et al. 2008). Above 13 km, we extrapolated the profile with an electronic noise of 1.38$\times$10$^{-18}$. See the text for further details.}
\label{cn2_domec}
\end{figure}

\begin{figure}
\begin{center}
\includegraphics[width=5cm,angle=-90]{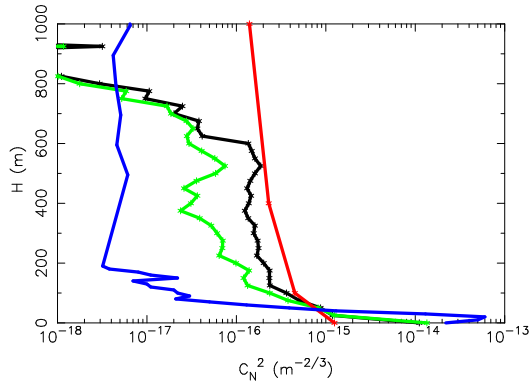}
\end{center}
\caption{Vertical distribution of the optical turbulence ($\CN2$ profiles). Blue line: measurements (radiosoundings) above Dome C in winter time (extracted from Trinquet et al. 2008). Red line: Hufnagel model describing how the turbulence typically decreases above the ground (Roddier 1981). Black line: measurements (Generalized Scidar) above Mt. Graham (extracted from Masciadri et al. 2008) in which the threshold of the instrument is considered ($\CN2$ = 10$^{-16}$). Green line: as the black line but we assume the optimistic case in which the absence of the signal corresponds to the absence of turbulence. The green line represents therefore what it would be if we would not be limited by the instrument sensitivity.}
\label{cn2_huf}
\end{figure}

A few years ago, when the ARENA coordination action started and astronomers showed their great interests for Dome C (and Antarctica), our believes was that, above 30 m from the ground (in which in this region the turbulence strength was even worse than above a mid-latitude site), the turbulence strength was so weak to let us think that Antarctica could reproduce on the Earth atmospherical conditions similar to what experienced in the space. Above the internal antarctic plateau the boundary layer appeared squeezed from the typical 1 km to $\sim$ 30 m with a median seeing in the free atmosphere $\varepsilon_{FA}$$=$ 0.27 arcsec, an isoplanatic angle $\thetaO$$= $5.7 arcsec and a wavefront coherence time $\tauO$$=$ 7.9 msec (Lawrence et al., 2004). Since there further site testing campaigns done with different instruments have been performed above Dome C (Trinquet et al. 2008, Aristidi et al. 2009). Which are at present the new insights we achieved on the turbulence characteristics in winter time in Antarctica with respect to those achieved so far above mid-latitude sites ? Let's start with the $\thetaO$, the most sensible indicator to quantity the turbulence in the high part of the atmosphere. 
The most recent statistical estimates done with a Generalized Scidar (GS) above Mt. Graham provide a median value of $\thetaO$$=$ 2.5 arcsec (Masciadri et al. 2008, 43 nights). This value is consistent with the median $\thetaO$$=$ 2.56 arcsec quantified with the same instrument on an even richer statistical sample ($\sim$ 100 nights) above the Teide (Tenerife) (Garcia-Lorenzo et al. 2009). With a different instrument (MASS), the isoplanatic angle above Mauna Kea on a statistic of three years of observations revealed to be $\thetaO$$=$ 2.6 arcsec (Schoeck et al. 2009). If we look at the median $\CN2$ profile measured in winter time above Dome C with balloons  (Fig.\ref{cn2_domec}) up to 13 km (where the balloons in general explode) and we extrapolate in the [13 - 20] km range the $\CN2$ with a value correspondent to the typical electronic noise, we obtain a median $\thetaO$$=$ 3.32 arcsec that is still better than the typical $\sim$ 2.5 arcsec measured at the mid-latitude sites but is seriously smaller than $\thetaO$$=$5.7 arcsec, that have been indicated during the preliminary site testing campaigns (Lawrence et al. 2004). We have therefore, in the most optimistic case, a gain of 33\% instead of the previously claimed 128\%. The impression is therefore that it is not really in the free atmosphere that we will have the greatest gain above Dome C. The most recent median values of $\tauO$ above Mt. Graham is 4.8 msec (Masciadri et al., 2008), above Teide is 4.97 msec (Garcia-Lorenzo 2009) compared to 3.7 msec (h$_{0}$= 8 m) and 5.2 msec (h$_{0}$= 30 m) (Trinquet et al. 2008) at Dome C. Also for $\tauO$ the gain achieved above Dome C with respect to mid-latitude sites is therefore seriously reduced with respect to what estimated a few years ago. If we look at the turbulence distribution in the first kilometer (Fig.\ref{cn2_huf}) measured above Dome C we see that the gain remains with respect to a mid-latitude site (Mt. Graham for example) even if, recently, it has been proved (Masciadri et al. 2008, Chun et al., 2009) that, even above a mid-latitude sites, the turbulence decreases in a much more sharper way that what supposed so far (see Hufnagel model - Fig.\ref{cn2_huf}). We conclude therefore that the challenge of Dome C for astronomical applications should be better express in the following way: 
\begin{itemize}
\item In the [30 m, 1 km] range the optical turbulence is substantially weaker than above mid-latitude sites. 
\item In the [0 m, 30 m] range the optical turbulence is substantially stronger than above mid-latitude sites. 
\end{itemize}
Does this condition get Dome C a peculiar site for some specific astronomical applications ? We think that this is the key question we have to answer to. The strength of the turbulence in the surface layer has to be carefully taken into account to specify the characteristics of AO systems in the case in which (as we will see later on) one intends to figure out a telescope with a primary mirror not completely free from the surface layer.

\section{Optical Turbulence characterization}

In this general panorama  in the context of the ForOT project a few years ago it has been undertaken an extended study of site assessment of Dome C (still on-going) that developed along three major axes of research: the characterization of the meteorological parameters characterization, the optical turbulence simulation with atmospherical mesoscale models and the performances of AO systems in such a regions. We report the status of art of results achieved putting the accent on those relevant for ARENA scientific goals.

\subsection{Meteorological parameters characterization}

The optical turbulence is just one of a set of parameters that can characterize the atmosphere for astronomical applications. Many of the classical meteorological parameters such as the wind speed, the temperature, the humidity, etc... are, by themselves, important elements to classify and characterize an astronomical site. They are also important indicators for the optical turbulence because its characteristics depend on most of these parameters. We refer the reader to a more detailed description of the results of this study to Hagelin et al. (this Conference Proceedings). We cite here the most important results relevant for the ARENA scientific goals:\newline
{\bf (1)} Studying the Richardson number at large spatial scales (method proposed by Geissler \& Masciadri 2006) we proved that the probability to trigger the optical turbulence in the free atmosphere above the Internal Antarctic Plateau is lower than above mid-latitude sites. If we limit the analysis to sites in Antarctica, Dome C showed the highest probability for turbulence triggering while South Pole the lowest one. \newline
{\bf (2)} Dome C is not necessarily the best site on the plateau. Above this site it has been observed, during the winter time, the largest median wind speed in the free atmosphere ($\sim$ 30 m$\cdot$s$^{-1}$ at 20 km) due to the polar vortex affecting the air circulation at synoptic scale as well as a strong wind shear in the first tens of meters near the surface (8 m$\cdot$s$^{-1}$ at 20 m). Such a strong wind shear combined with a near ground  thermal stratification is the principal source for the triggering of the optical turbulence in this thin vertical slab. \newline
{\bf (3)} Dome A showed the highest thermal stability near the surface as well as the highest wind speed shear. This let us expect at Dome A a surface layer of comparable size or even thinner that that developed at Dome C but probably with a larger turbulence strength inside the thin surface layer. \newline
{\bf (4)} A detailed study (Lascaux et al., 2008, Fig. 2) of the median wind speed in the free atmosphere supported our thesis claiming that the farthest from the center of the polar vortex is the site location, the larger is the wind speed strength and shear above 10 km. This evidence suggest an objective criterium for the selection of sites with the lowest wind speed in the high part of the atmosphere. This might be particularly important to maintain large values of $\tauO$.\newline
{\bf (5)} The European Center for Medium Weather Forecasts (ECMWF) analyses can hardly be used to discriminate site atmospheric characteristics near the ground and this justifies the use of atmospherical mesoscale models.

\subsection{Optical turbulence reconstructed with atmospheric mesoscale models}

A mesoscale model (Meso-Nh) has been used to perform a study on the ability of such a model in reconstructing the most important classical meteorological parameters (T, wind speed, ect...) and the parameters characterizing the optical turbulence (the optical turbulence vertical distribution in different vertical slabs within 20 km and the thickness of the surface layer). The use of atmospherical mesoscale models can play a key role in the characterization of the turbulence. We refer the reader for an exhaustive presentation of these challenges to Masciadri (2006). We summarize here the most important results obtained so far and we refer to Lascaux et al. (2009) for the details of this study.

Two different model configurations have been used to perform a statistical comparison analysis between measurements and simulations. A monomodel configuration (that we will refer as 'low resolution') in which the whole Antarctic continent is covered with 60$\times$60 grid points and a horizontal resolution $\Delta$X=100 km and a grid-nesting configuration (that we will refer as 'high resolution') made by three models that refine the resolution around the selected site achieving a maximum resolution of $\Delta$X= 1km on a domain of 80$\times$80 grid points.\newline
{\bf (1)} Using a statistic sample of 47 nights it has been possible to prove that Meso-Nh reconstructs as well as the ECMWF analyses the temperature and the wind speed in the h $\ge$ 1 km region. However, Meso-Nh better reconstructs the temperature and the wind speed near the ground (h $\le$ 1 km) than the ECMWF analyses. It has been possible to demonstrate that Meso-Nh is 2-2.5 times more accurate than the ECMWF analyses ($\Delta$T$_{ecmwf,obs}$$=$3.74 K, $\Delta$T$_{mnh,obs}$$=$1.6 K) for the temperature and 60 times more accurate for the wind speed ($\Delta$V$_{ecmwf,obs}$$=$ 2.49 ms$^{-1}$, $\Delta$V$_{mnh,obs}$$=$ 0.04 2.49 ms$^{-1}$). \newline
{\bf (2)} The simulated $\CN2$ profiles of all the 15 nights in winter 2005 [21 June - 21 September] for which measurements obtained with balloons equipped with thermal micro-sensors are available (see Trinquet et al. 2008) have been compared to the observed $\CN2$ measurements. 
Defining the thickness of the surface layer as Eq.(1) (Lascaux et al. 2009), we obtained a mean value of the observed surface layer thickness h$_{sl,obs}$$=$35.3$\pm$19.9 m versus a mean value of the simulated surface layer h$_{sl,mnh-high}$$=$48.9$\pm$29.3 m obtained with the Meso-Nh model with the high resolution. It has been possible to conclude that the simulations well match the observation because the $\Delta$h $\sim$ 13 m between the observations and simulations means is within the dispersion of measurements $\sigma$$=$19.9 m. \newline
{\bf (3)} It has been observed that the simulations obtained with the low resolution of $\Delta$X$=$100 km provide not enough accurate results (the difference between the observed and simulated means is larger than the dispersion). \newline
{\bf (4)} The ability of the model in reconstructing the vertical distribution of the optical turbulence ($\CN2$ profiles) has been tested calculating the two parameters: \newline

\begin{equation}
\varepsilon_{TOT}=5.41 \cdot  \lambda^{-1/5} \cdot \left( \int_{8m}^{h_{top}} C_N^2(h) \cdot dh \right) ^{3/5}
\label{seetot}
\end{equation}

\begin{equation}
\varepsilon_{FA}=5.41 \cdot \lambda^{-1/5} \cdot \left( \int_{h_{sl}}^{h_{top}} C_N^2(h) \cdot dh \right) ^{3/5}
\label{seefa}
\end{equation}

where h$_{top}$ $=$ 13 km, i.e. the height at which balloons explode. Table \ref{see_stat} shows the results obtained:\newline
\begin{table}
\begin{tabular}{ccccccccc}
\hline
STAT & Obs & & & Model  & & &Model   &  \\
 & & & & $\Delta$X= 1 km &  & & $\Delta$X= 100 km & \\
 & (") & (") & & (") & (") & &  (") & (") \\
 & $\varepsilon_{TOT}$ & $\varepsilon_{FA}$ & &  $\varepsilon_{TOT}$& $\varepsilon_{FA}$& &$\varepsilon_{TOT}$ & $\varepsilon_{FA}$ \\
\hline
Median & 1.6 & 0.3 & & 2.29 &0.35 & & 3.58 &0.42 \\
$\sigma$ & 0.7 &0.7  & & 1.46 &0.92 & & 1.64 & 1.07 \\
\hline
\end{tabular}
\caption{Statistics (median and $\sigma$) of the observed and simulated $\varepsilon_{TOT}$ and  $\varepsilon_{FA}$ related to the sample of 15 nights in winter time 2005 (Lascaux et al. 2009). h$_{sl,obs}$$=$35.3 m, h$_{sl,mnh-high}$$=$48.9 m, h$_{sl,mnh-low}$$=$65.9 m.}
\label{see_stat}
\end{table}

The high resolution shows, again more accurate results than the low resolution; we analyze therefore the best results obtained with this configuration. The simulated median seeing in the free atmosphere ($\varepsilon_{FA}$ $=$ 0.35 arcsec) is well correlated (within $\sigma$) with the observed one ($\varepsilon_{FA}$ $=$ 0.3 arcsec). The simulated total seeing is however overestimated with respect to the observed one. The excess of turbulence reconstructed by the model is located in the surface layer. Studies aiming to correct this trend are on-going.\newline
{\bf (5)} For the first time the $\CN2$ profiles have been simulated with a mesoscale model above Dome C up to 20 km and it has been shown (Lascaux et al. 2009, Fig. 8) that the model temporal variability is present even within a small dynamic. \newline

In the context of this extended study that aims to apply the mesoscale model to other locations above the antarctic plateau, it is worth to briefly refer to a study undertaken with some collaborators (Le Moigne et al. 2008) aiming to optimize the surface scheme of Meso-Nh, the so called ISBA (Interaction Soil Biosphere Atmosphere) scheme. It is indeed obvious that the most critical part of an atmospherical model for this kind of simulations if the scheme that controls the air/ground turbulent fluxes budget. Our ability in well reconstructing the surface temperature T$_{s}$ is related to the ability in well reconstructing the sensible heat flux H that is responsible of the buoyancy-driven turbulence in the surface layer. The Meso-Nh surface scheme is based on the force-restore method which consists on two equations that control the temporal evolution of the surface temperature T$_{s}$ and the deep temperature T$_{2}$. The equation of the temporal evolution of the deep temperature has been modified adding a term depending on two free parameters: a climatological temperature T$_{c}$ and a a relaxation term called $\gamma$. The two free parameters are fixed forcing the system of two equations all along one year by the radiative measurements of the solar direct radiation and the long-wavelength radiation\footnote{Kindly provided by S. Argentini.} and the (temperature T, wind speed V, pressure p and specific humidity q)\footnote{Kindly provided by A. Pellegrini.} of the air above the ground. The simulated T$_{s}$ and T$_{2}$ are compared to the observed one (i.e. the temperature measured at - 5 cm and -30 cm from the ground)\footnote{Kindly provided by S. Argentini.} up to minimize the dispersion. Such a study permitted us to optimize the surface scheme ISBA for applications of the Meso-Nh model to polar conditions.

\subsection{Ground Layer Adaptive Optics performances above Dome C}

\begin{figure}
\begin{center}
\includegraphics[width=5.5cm]{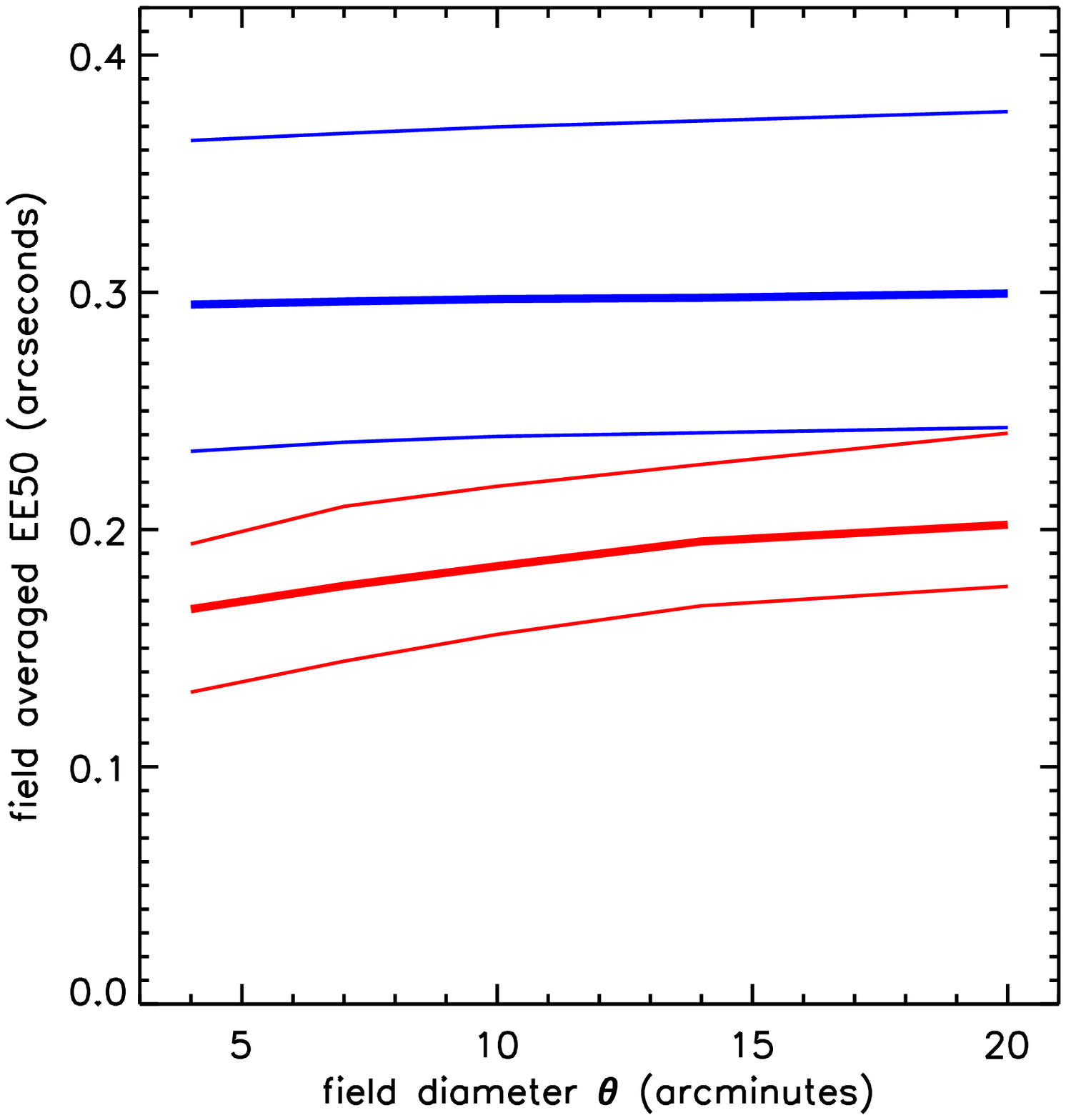}
\includegraphics[width=5.5cm]{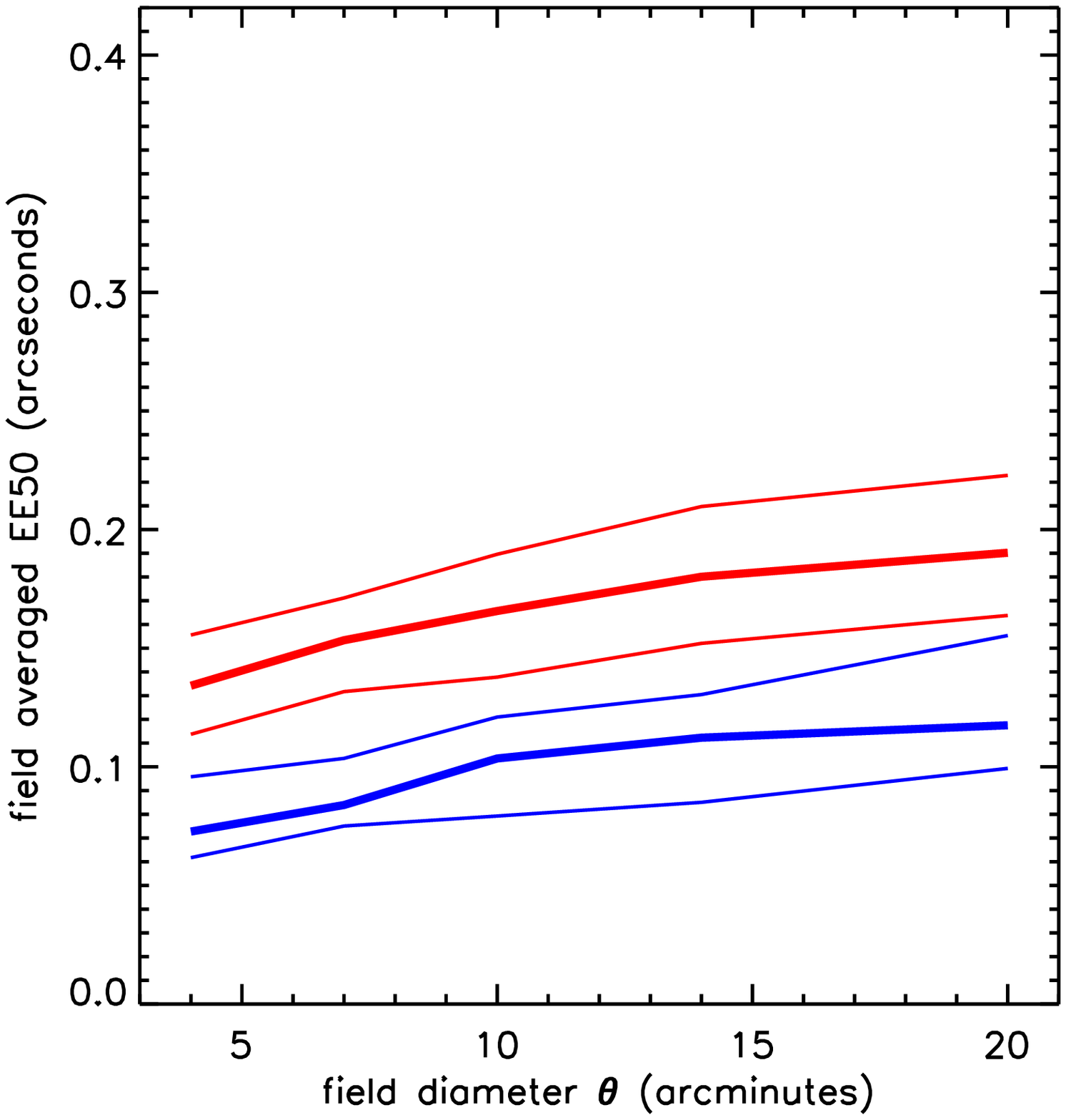}
\end{center}
\caption{EE50 i.e. angular size (") in which 50\% of the energy of the PSF is included versus the FOV. Red line: result obtained with the turbulence distribution above Mt. Graham. Blue line: Dome C. Thick lines are the median value, thin lines the quartiles. Left: the pitch size $\Delta$X=0.5 m. Right: the pitch size $\Delta$X=0.1 m. Specifications: J band, D= 8 m and 4 guide stars.}
\label{glao}
\end{figure}

As described in Section \ref{intro} the real challenge of Dome C for astronomical applications in the near-infrared is the wide field because the optical turbulence is concentrated in a thin surface layer in this region. The wider the angle the shorter is the depth of field on which the GLAO system can correct the turbulence efficiently. We can envisage two solutions: {\bf (1)} a telescope whose primary mirror is located above the surface layer (i.e. 50-60 m including the statistical dispersion). {\bf (2)} a telescope embedded in the strong surface turbulent layer equipped with a GLAO system conceived to correct  the surface layer and for which the final performances should be better than what achievable at mid-latitude sites. But...Is there any turbulence or technical constraints in this second hypothesis ? The wavefront correction stops typically on a spatial scale roughly equivalent to the r$_{0}$ for the wavelength we are considering and the correction performances strongly depends on the pitch size of the AO system. Figure \ref{glao} shows the results of simulations (Stoesz et al. 2008) obtained for a GLAO system conceived for a 8 m telescope equipped with 4 guide stars and different pitch size $\Delta$X in J band. The vertical turbulence distribution of Dome C (Trinquet et al. 2008) and Mt. Graham (Masciadri et al. 2008) have been considered. Mt. Graham is taken as representative of  a mid-latitude site. We can observe that, with $\Delta$X = 0.5 m (Fig.\ref{glao}-left side) the EE50 (i.e. angular size (") in which 50\% of the energy of the PSF is included versus the FOV) obtained above Dome C is larger than what achievable above Mt. Graham. To invert the tendency, i.e. to obtain a smaller angular size it is necessary to consider a smaller pitch size ($\Delta$X) that is we have to consider more sophisticated AO systems that correct higher orders of the wavefront perturbations (Fig.\ref{glao}-right side). For this set of simulations Dome C starts to be competitive for $\Delta$X = 0.38 m but to obtain a more accurate value of this threshold one should do Montecarlo simulations. It is however evident that for small telescopes (2 m class telescopes) it is relatively easy to figure out a standard GLAO system with a reasonable number of actuators that fit the constraint of a small pitch size. For a a larger telescope (8-10 m class telescopes or even more), on the contrary, we need to envisage a high order AO system to be competitive with an equivalent facilities at a mid-latitude site.

\section{Conclusion}

In this contribution we present the most important results/conclusions achieved so far by our group in a comprehensive study addressing the optical turbulence characterization above Dome C for astronomical applications. We answered to a set of open questions that might support the main ARENA decisions. \newline
{\bf (1)} We provided a wind speed site rank for h $>$ 1 km showing that Dome C has the strongest wind speed shear and strength in the high part of the atmosphere in winter time, South Pole the weakest one (Hagelin et al. 2008). This feature is probably the main cause for the $\tauO$ decreasing in this season and for the increasing of the probability to trigger local turbulence in the high atmosphere. \newline
{\bf (2)} Studying the inverse of the Richardson number (1/R$_{i}$) at large spatial scales we calculated the site rank (for h $>$ 1 km) of the probability to trigger thermodynamic instability (i.e. optical turbulence). Dome C shows the greatest probability to trigger optical turbulence in the free atmosphere particularly in winter time (Hagelin et al. 2008) but all the antarctic sites show a smaller probability to trigger turbulence in the free atmosphere than above mid-latitude sites . The development of local turbulence in the high atmosphere is certainly the cause of the smaller $\thetaO$ in the winter time. \newline
{\bf (3)} In the context of modeling we proved the necessity of mesoscale models for the characterization of the thermodynamic stability of the surface layer (the ECMWF analyses are not enough accurate) (Hagelin et al. 2008).\newline
{\bf (4)} We statistically demonstrated that Meso-Nh provides a more accurate description of the temperature (T) and the wind speed (V) near the ground with respect to ECMWF analyses (Lascaux et al. 2009). This is promising for a good reconstruction of the optical turbulence done by the mesoscale models. \newline
{\bf (5)} The typical thickness of the surface layer (h$_{sl}$) and the seeing developed above h$_{sl}$ ($\varepsilon_{FA}$) are statistically well reconstructed by the Meso-Nh mesoscale model and they are well correlated with measurements. The model, at present, overestimates the strength of the turbulence in the surface layer, further studies are on-going to correct this trend. (Lascaux et al. 2009). \newline
{\bf (6)} An optimization of the Meso-Nh model surface scheme (ISBA) in antarctic conditions has been performed (Le Moigne et al. 2008) permitting to improve the model reconstruction of the temperature underground and, as a consequence, the sensible heat flux ground/air, the principal driver of the dynamical turbulence in the antarctic surface layer. \newline
{\bf (7)} The most interesting challenge for astronomical applications in the [J-K] band is the wide-field imaging equipped with GLAO systems. \newline
{\bf (8)} We demonstrated that, in a GLAO system, the pitch size projected on the pupil of the telescope should smaller than 0.5 m in J band to achieve, within 20', better performances  that what is achievable with an equivalent system above mid-latitude sites. For 2 m class telescopes this condition is easily obtainable. For 8-10 m class telescopes a high-or-de AO correction and a large number of actuators is required to perform astronomical observations competitive with respect to what achievable at mid-latitude sites (Stoesz et al. 2008) .\newline
\section{Acknowledgments} 
This study has been funded by the Marie Curie Excellence Grant (ForOT) 
MEXT-CT-2005-023878 - FP6 Program. ECMWF products are extracted from the MARS-catalog, $http://www.ecmwf.int$. 
Meteorological parameters measurements from the Progetto di Ricerca 'Osservatorio Meteo Climatologico' of the Programma Nazionale di Ricerche in Antartide (PNRA), $http://www.climantartide.it$ and from the AMRC (Antarctic Meteorological Research Center, University of Wisconsin, Madison $ftp://amrc.ssec.wisc.edu/pub/southpole/radiosonde$). We acknowledge the ISAC/CNR (Italy) group (S. Argentini) for providing in-situ measurements (radiative fluxes and underground temperature).
\newline

\end{document}